\documentclass[aps,prc,letterpaper,11pt,twoside,tightenlines,nofootinbib,
showpacs,preprint]{revtex4}
\usepackage{graphicx}
\usepackage{epsfig,float}
\usepackage[sort&compress]{natbib}
\usepackage{subfigure}
\usepackage{amsmath}
\usepackage{amsfonts}
\usepackage{cancel}
\usepackage{lmodern,dsfont}
\usepackage{amssymb}
\usepackage{hyperref}
\begin{document}
\arraycolsep1.5pt
\newcommand{\Ima}{\textrm{Im}}
\newcommand{\Rea}{\textrm{Re}}
\newcommand{\mev}{\textrm{ MeV}}
\newcommand{\gev}{\textrm{ GeV}}
\newcommand{\dtres}{d^{\hspace{0.1mm} 3}\hspace{-0.5mm}}
\newcommand{\rts}{ \sqrt s}
\newcommand{\non}{\nonumber \\[2mm]}
\newcommand{\eps}{\epsilon}
\newcommand{\half}{\frac{1}{2}}
\newcommand{\thalf}{\textstyle \frac{1}{2}}
\newcommand{\Nmass}{M_{N}} % mass of nucleon
\newcommand{\delmass}{M_{\Delta}} % mass of delta
\newcommand{\pimass}{\mu}  % mass of pion 
\newcommand{\rhomass}{m_\rho} % mass of rho
\newcommand{\piNN}{f}      % coupling of pi NN
\newcommand{\rhocoup}{g_\rho} % universal coupling to rho
\newcommand{\fpi}{f_\pi} % pion decay constant fpi
\newcommand{\f}{f} % pion decay constant fpi
\newcommand{\nucfld}{\psi_N} % nucleon field
\newcommand{\delfld}{\psi_\Delta} % delta field
\newcommand{\fpiNN}{f_{\pi N N}} % coupling of pi N N 
\newcommand{\fpiND}{f_{\pi N \Delta}} % coupling of pi N delta 
\newcommand{\GMquark}{G^M_{(q)}} % magnetic coupling for quark 
\newcommand{\vecpi}{\vec \pi}
\newcommand{\vectau}{\vec \tau}
\newcommand{\vecrho}{\vec \rho}
\newcommand{\delmu}{\partial_\mu}
\newcommand{\delMu}{\partial^\mu}
\newcommand{\nn}{\nonumber}
\newcommand{\bi}{\bibitem}
\newcommand{\vs}{\vspace{-0.20cm}}
\newcommand{\be}{\begin{equation}}
\newcommand{\ee}{\end{equation}}
\newcommand{\ba}{\begin{eqnarray}}
\newcommand{\ea}{\end{eqnarray}}
\newcommand{\ropi}{$\rho \rightarrow \pi^{0} \pi^{0}
\gamma$ }
\newcommand{\roeta}{$\rho \rightarrow \pi^{0} \eta
\gamma$ }
\newcommand{\omepi}{$\omega \rightarrow \pi^{0} \pi^{0}
\gamma$ }
\newcommand{\omeeta}{$\omega \rightarrow \pi^{0} \eta
\gamma$ }
\newcommand{\ul}{\underline}
\newcommand{\del}{\partial}
\newcommand{\rth}{\frac{1}{\sqrt{3}}}
\newcommand{\rsix}{\frac{1}{\sqrt{6}}}
\newcommand{\sq}{\sqrt}
\newcommand{\fr}{\frac}
\newcommand{\pr}{^\prime}
\newcommand{\ov}{\overline}
\newcommand{\Gm}{\Gamma}
\newcommand{\rw}{\rightarrow}
\newcommand{\rgl}{\rangle}
\newcommand{\De}{\Delta}
\newcommand{\Dp}{\Delta^+}
\newcommand{\Dm}{\Delta^-}
\newcommand{\Dz}{\Delta^0}
\newcommand{\Dpp}{\Delta^{++}}
\newcommand{\Sg}{\Sigma^*}
\newcommand{\Sp}{\Sigma^{*+}}
\newcommand{\Sm}{\Sigma^{*-}}
\newcommand{\Sz}{\Sigma^{*0}}
\newcommand{\X}{\Xi^*}
\newcommand{\Xm}{\Xi^{*-}}
\newcommand{\Xz}{\Xi^{*0}}
\newcommand{\Om}{\Omega}
\newcommand{\Omm}{\Omega^-}
\newcommand{\kp}{K^+}
\newcommand{\kz}{K^0}
\newcommand{\pip}{\pi^+}
\newcommand{\pim}{\pi^-}
\newcommand{\piz}{\pi^0}
\newcommand{\et}{\eta}
\newcommand{\kb}{\ov K}
\newcommand{\km}{K^-}
\newcommand{\kbz}{\ov K^0}
\newcommand{\ksb}{\ov {K^*}}

\def\tstrut{\vrule height2.5ex depth0pt width0pt} % used in tables
\def\jtstrut{\vrule height5ex depth0pt width0pt} % used in tables

\title{The role of vector-baryon channels and resonances in the $\gamma p \to K^0 \Sigma^+$ and $\gamma n \to K^0 \Sigma^0$ reactions near the $K^* \Lambda$ threshold.
}

\author{A. Ramos$^1$ and E. Oset$^2$}
\affiliation{
$^1$ Departament d'Estructura i Constituents de la Materia and Institut de Ciencies del Cosmos, Universitat de
Barcelona , Mart\'{\i} i Franqu\`es 1, 08028 Barcelona, Spain \\
$^{2}$Departamento de F\'{\i}sica Te\'orica, Universidad de Valencia and
IFIC, Centro Mixto Universidad de 
Valencia-CSIC,
Institutos de Investigaci\'on de Paterna, Aptdo. 22085, 46071 Valencia,
Spain
}

\date{\today}

\begin{abstract}
We have studied the $\gamma p \to K^0 \Sigma^+$ reaction in the energy region
around the $K^* \Lambda$ and $K^* \Sigma$ thresholds, where the CBELSA/TAPS
cross section shows a sudden drop and the differential cross section
experiences a transition from a forward-peaked distribution to a flat
one. Our coupled channel model incorporates the dynamics of the vector
meson-baryon interaction which is obtained from the hidden gauge formalism. We
find that the cross section in this energy region results from a delicate 
interference between amplitudes having $K^* \Lambda$
and $K^*\Sigma$ intermediate states. The sharp downfall is dictated by the presence of a nearby $N^*$ resonance produced by our model, a feature
that we have employed to predict its properties. We also show results for the
complementary $\gamma n \to K^0 \Sigma^0$ reaction, the measurement of which
would test the mechanism proposed in this work.
\end{abstract}

\maketitle

\section{Introduction}
\label{Intro} 
The recent work reported by the CBELSA/TAPS collaboration \cite{schmieden} puts a
challenge to the ordinary models of photoproduction of mesons. The reaction is
$\gamma p \to K^0 \Sigma^+$, which exhibits a peak in the cross section around 
$\sqrt s =1900$~MeV followed by a fast downfall around $\sqrt s =2000$~MeV. Most
remarkable, the differential cross section is flat close to threshold, becomes
forward peaked close to the energy where the cross section has a maximum but, up to the resolution of the experiment,
appears again isotropic in the region where the total cross section is small and
nearly constant, from $\sqrt s =2000$~MeV to 
$\sqrt s =2200$~MeV. The experiment complements and improves earlier measurements of
Crystal Barrel \cite{cb} and SAPHIR \cite{saphir}. As shown in \cite{schmieden},
sophisticated models of $K$ photoproduction like K-MAID \cite{maid} and SAID
\cite{said} grossly fail to reproduce the experimental features, even when
changes are made to adapt the models to this particular reaction. The same fate
is shown to follow for the models
\cite{Anisovich:2005tf,Sarantsev:2005tg,Usov:2005wy}, as
discussed in \cite{saphir}. 
 
   In this work we present a theoretical approach that gives an
explanation to the features observed in the $\gamma p \to K^0 \Sigma^+$ reaction around
$\sqrt s =2000$~MeV. The experimental paper 
\cite{schmieden} hints to 
some mechanisms involving  vector meson-baryon channels,
since the prominent
feature discussed above occurs in between the $K^* \Lambda$ and $K^* \Sigma$
thresholds, at $2010$~MeV and $2087$~MeV, respectively. 
Our model implements the vector-baryon interaction for vectors of the nonet with the octet of baryons
obtained in \cite{angelsvec}, using the local hidden gauge
Lagrangians \cite{hidden1,hidden2,hidden4} and coupled channels in an unitary
approach. This vector-baryon interaction leads to the dynamical generation of
$1/2^-$ and $3/2^-$ resonances, degenerate in spin-parity, one of which appears
around 1970~MeV and couples to $\rho N$, $\omega N$, $\phi N$ but mostly to $K^*
\Lambda$ and $K^* \Sigma$. The fact that this resonance appears close to the location of the downfall of the cross section suggests that any realistic theoretical scheme trying to reproduce this problem should consider the explicit incorporation of these
channels and their interaction, as done in the present work.  We show that the interference of the  $K^*
\Lambda$ and $K^* \Sigma$ channels, magnified by the presence of the resonance,
is essential for reproducing the behavior of the $\gamma p \to K^0 \Sigma^+$
cross section  around $2000$~MeV. We also give predictions for the $\gamma n \to
K^0 \Sigma^0$ reaction, which has a quite different interference pattern and
shows a peak in the energy region where the $\gamma p \to K^0 \Sigma^+$ has the
downfall. A measurement of the neutral reaction could then bring further light
into the physics hidden in these processes.

\section{Formalism}
\label{Formal}

\begin{figure}[htb]
\begin{center}
\includegraphics[width=0.4\textwidth]{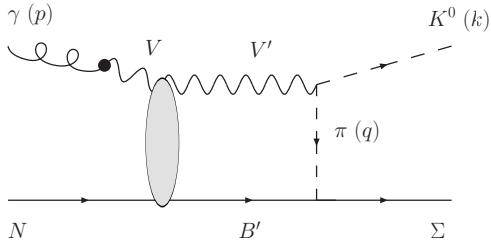}
\caption{Mechanism for the photoproduction reaction $\gamma N \to K^0\Sigma$.
The symbol $V$ stands
for the $\rho^0$, $\omega$ and $\phi$ mesons, while $V^\prime B^\prime$ denotes
the intermediate channel, which can be
$K^{*+}\Lambda$, $K^{*+}\Sigma^0$ or
$K^{*0}\Sigma^+$,
in the case of $\gamma p \to K^0\Sigma^+$, 
and $K^{*+}\Sigma^-$  or
$K^{*0}\Lambda$,
in the case of $\gamma n \to K^0\Sigma^0$, the $K^{*0}\Sigma^0$ one not
contributing in the later reaction due to the zero value of the
$\pi^0\Sigma^0\Sigma^0$ coupling at the Yukawa vertex. }
\label{fig:diag}
\end{center}
\end{figure}

The hidden gauge approach incorporates automatically
vector meson dominance \cite{sakurai}, converting the photon into a vector meson, 
which later on interacts with the other hadrons. Our basic mechanisms for the $\gamma N \to K^* \Sigma$ reaction are depicted in Fig. \ref{fig:diag}, where we can see the
photon conversion into $\rho^0, \omega, \phi$, followed by the $\rho N, \omega
N,
\phi N$
interaction leading to the relevant vector-baryon ($V^\prime B^\prime$)
channels, which are of $K^*\Lambda$ or  $K^* \Sigma$  type,
% ($K^{*+} \Lambda$,
%$K^{*+}
%\Sigma^0$, $K^{*0}\Sigma^+$ for $\gamma p \to K^0\Sigma^+$, and $K^{*0}
%\Lambda$, $K^{*+}
%\Sigma^-$ for $\gamma n \to K^0\Sigma^0$), 
since those are
the ones to which the resonance around 1970~MeV couples most strongly
according to the model of Ref.~\cite{angelsvec}. 
Finally, the
intermediate $K^*  \Lambda$ or $K^* \Sigma$ states get converted via pion
exchange to the final $K^{0} \Sigma^+$, in the case of $\gamma p$, or $K^{0}
\Sigma^0$, in the case of $\gamma n$. We
take the vector-baryon amplitudes $V N \to V^\prime B^\prime$ from the work
of Ref.~\cite{angelsvec}.
Since in the final state we have a pseudoscalar meson and a baryon, it
would be most appropriate to work with a model space that contains
both the pseudoscalar-baryon and the vector-baryon channels, as done in
\cite{javier,kanchan2,kanchan3}. Yet, at the energy that we are concerned we can
neglect the interaction of the pseudoscalar-baryon channels since they do not
produce any resonance around this region \cite{inoue}, and then consider
just the interaction of the vector-baryon channels plus the mechanism
responsible for the transition from vector-baryon to the final
pseudoscalar-baryon that we take from \cite{javier}.  The formalism of
Ref.~\cite{angelsvec} was developed under the assumption that the momenta of the
external mesons was small. In the present work, the momentum of the photon and,
hence, that of the virtual $\rho^0$, $\omega$ or $\phi$ mesons, is not small,
but some of the simplifying assumptions of the model of Ref.~\cite{angelsvec}
are still valid or have a limited influence. On the one hand, neglecting the
zeroth component of the polarization vector, $\epsilon^0$, is still possible for
these mesons since they acquire the polarization vector of the photon, which is
of transverse nature. On the other hand, we have estimated that the size of the
linear momentum terms
neglected here to be about 15\% of the dominant contributions to
the $VB \to V^\prime B^\prime$ interaction. This observation can justify a
posteriori why the small three-momentum approximation applied in the radiative
decays of vector-vector molecules \cite{raquel,yamagata,branz} led to such good
results in spite of the finite momentum of the emitted photons.
Further
details on the interaction of vectors with mesons and baryons can be seen in the
review \cite{review} and in Ref.~\cite{hidekoroca}.

 The
photon-vector conversion Lagrangian is given by
\begin{equation}
 {\cal L}_{\gamma V}=- M_V^2 \frac{e}{g} A_\mu \langle V^\mu Q \rangle \ ,
\end{equation}
with $Q=diag(2,-1,-1)/3$ and $A_\mu$ being the photon field. The charge of the
electron, $e$, is negative and normalized as $e^2/4\pi=1/137$.

The $V^\prime PP$ vertex is obtained from
 \be
{\cal L}_{VPP}= -ig \langle [P,\partial_{\mu}P]V^{\mu}\rangle \ ,
\label{lagrVpp}
\ee
with the coupling of the theory given by $g=\frac{M_V}{2f}$
where $f=93$~MeV is the pion decay constant and $M_V$ the vector-meson mass. The
magnitude $V_\mu$ is the SU(3) 
matrix of the vectors of the $\rho$ nonet and $P$ stands for the matrix of the
pseudoscalar mesons of the $\pi$.

 Finally, the Yukawa vertex
is described by the Lagrangian:
\begin{equation}
 {\cal L}_{PBB}=\frac{1}{2}(D+F) \langle \bar{B}\gamma^\mu\gamma^5 u_\mu
B\rangle +
\frac{1}{2}(D-F) \langle \bar{B}\gamma^\mu\gamma^5 B u_\mu
\rangle \ ,
\end{equation}
where the term
$\gamma^\mu\gamma^5 u_\mu$ is taken in its non-relativistic form:
\begin{equation}
 \gamma^\mu\gamma^5 u_\mu \to \frac{\sqrt{2}}{f} \sigma^i\partial_i\phi \ ,
~~~~~~~i=1,2,3 \ , 
\end{equation}
$B$ is the matrix representing the baryon octet, and $D = 0.795$,
$F = 0.465$ are taken from \cite{Borasoy:1998pe}.

The amplitude corresponding to the mechanism in Fig.~\ref{fig:diag} is given by:
\begin{eqnarray}
 -i t^{\pi {\rm-pole}}_{\gamma N \to K^0 \Sigma} = && e
\sum_{V=\rho^0,\omega,\phi}
{\cal C}_{\gamma V} \sum_{V^\prime B^\prime} t_{V N \to V^\prime B^\prime}\,\, i
\int
\frac{d^4q}{(2\pi)^4} \frac{1}{(q+k)^2-M_{V^\prime}^2+ i\varepsilon}
 \frac{1}{q^2-m_\pi^2+ i\varepsilon} \nonumber
\\ &&
\frac{M_{B^\prime}}{E_{B^\prime}}\frac{1}{P^0-q^0-k^0-E_{B^\prime}(\vec{q}+\vec{
k}\,) + i\varepsilon} (\vec{q}-\vec{k}\,) \vec{\epsilon}_\gamma \,
\vec{\sigma}\vec{q}\,\, V_{Y,B^\prime} F(q) \ ,
\label{eq:ampl}
\end{eqnarray}
where
\begin{equation}
 {\cal C}_{\gamma V}=\begin{cases}
                             \frac{1}{\sqrt{2}} & \mbox{for }
V=\rho^0  \\ \frac{1}{3\sqrt{2}} & \mbox{for }
V=\omega \\
-\frac{1}{3} & \mbox{for }
V=\phi
                            \end{cases} \ ,
\end{equation}
and
\begin{equation}
 V_{\Sigma^+,B^\prime} = \begin{cases} 
                          \frac{2D}{2f\sqrt{3}} & \mbox{for }
V^\prime B^\prime=K^{*+} \Lambda \\
-\frac{2F}{2f} & \mbox{for } V^\prime B^\prime=K^{*+} \Sigma^0 \\
\frac{2F}{2f}\left(-\frac{1}{\sqrt{2}}\right) & \mbox{for } V^\prime
B^\prime=K^{*0} \Sigma^+
                         \end{cases} \ ,
\end{equation}
in the case of $\gamma p \to K^0 \Sigma^+$, while
\begin{equation}
 V_{\Sigma^0,B^\prime} = \begin{cases}
\frac{2F}{2f} & \mbox{for } V^\prime B^\prime=K^{*+} \Sigma^- \\
\frac{2D}{2f\sqrt{3}}\left(-\frac{1}{\sqrt{2}}\right) & \mbox{for } V^\prime
B^\prime=K^{*0} \Lambda
                         \end{cases} \ ,
\end{equation}
in the case of $\gamma n \to K^0 \Sigma^0$. Note that the factor $-1/\sqrt{2}$
appearing in $V_{Y,B^\prime}$ when $V^\prime=K^{*0}$ accounts for the
relation between the neutral meson coupling,
$K^{*0} \to \pi^0 K^0$,
to the charged meson one, $K^{*+} \to \pi^+ K^0$, in the $V^\prime PP$ vertex.
For the form factor $F(q)$ we take a typical Yukawa static shape,
$\Lambda^2/(\Lambda^2 + \vec{q}\,^2)$ with a cut-off $\Lambda=850$~MeV, a value
that has been adjusted to reproduce the size of the experimental $\gamma p \to
K^0 \Sigma^+$ cross section. One could also argue that there can be an extra form factor factor at the meson $VPP$ vertex. In such a case, one should consider our prescription as an empirical way of accounting for these combined finite-size effects. 

\begin{figure}[htb]
\begin{center}
\includegraphics[width=0.4\textwidth]{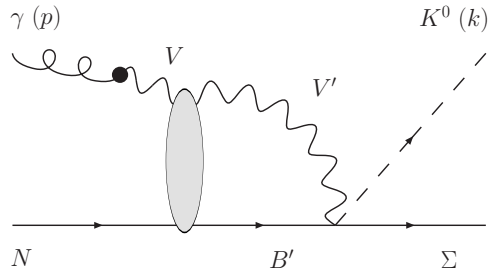}
\caption{Kroll-Ruderman contact term to be added to the mechanisms of
Fig.~\protect\ref{fig:diag} to preserve gauge-invariance in the photoproduction
reactions $\gamma N \to K^0\Sigma$. }
\label{fig:KR}
\end{center}
\end{figure}

In order to implement gauge invariance,  the amplitudes of Eq.~(\ref{eq:ampl})
must be complemented by the corresponding Kroll-Ruderman
contact term \cite{javier,kanchan2,kanchan3}, displayed in Fig.~\ref{fig:KR}
and given by
\begin{eqnarray}
 -i t^{KR}_{\gamma N \to K^0 \Sigma} = && e \sum_{V=\rho^0,\omega,\phi}
{\cal C}_{\gamma V} \sum_{V^\prime B^\prime} t_{V N \to V^\prime B^\prime}\, i
\int
\frac{d^4q}{(2\pi)^4} \frac{1}{(q+k)^2-M_{V^\prime}^2+ i\varepsilon} \nonumber
\\ &&
\frac{M_{B^\prime}}{E_{B^\prime}}\frac{1}{P^0-q^0-k^0-E_{B^\prime}(\vec{q}+\vec{
k}\,) + i\varepsilon} \, \vec{\sigma} \vec{\epsilon}_\gamma \,
 V_{Y,B^\prime} \, F(q) \ .
\label{eq:KR}
\end{eqnarray}
After performing the $q^0$ integration analytically,  Eqs.~(\ref{eq:ampl}) and
(\ref{eq:KR}) both develop two structures, $\vec{\sigma} \vec{k} \,
\vec{\epsilon}_\gamma \vec{k}$ and $\vec{\sigma} \vec{\epsilon}_\gamma$. Adding all contributions together, we obtain
\begin{equation}
-i t_{\gamma N \to K^0 \Sigma}= A  \vec{\sigma} \vec{k}\,\vec{\epsilon}_\gamma
\vec{k} +  B \vec{\sigma}  \vec{\epsilon}_\gamma \ ,
\label{eq:ampltot}
\end{equation}
where
\begin{equation}
A=e \sum_{V=\rho^0,\omega,\phi}
{\cal C}_{\gamma V} \sum_{V^\prime B^\prime} t_{V N \to V^\prime B^\prime} \, 
V_{Y,B^\prime}\, \tilde{G}^{(1)}_{V^\prime B^\prime}  \ ,
\label{eq:A}
\end{equation}
and
\begin{equation}
B=e \sum_{V=\rho^0,\omega,\phi}
{\cal C}_{\gamma V} \sum_{V^\prime B^\prime} t_{V N \to V^\prime B^\prime} \, 
V_{Y,B^\prime} \, \left(\tilde{G}^{(2)}_{V^\prime B^\prime} + {G}_{V^\prime
B^\prime}\right) \ .
\label{eq:B}
\end{equation}
The loop functions $\tilde{G}^{(1)}_{V^\prime B^\prime} $, $\tilde{G}^{(2)}_{V^\prime B^\prime} $ and ${G}_{V^\prime B^\prime}$ are given by 
\begin{eqnarray}
\tilde{G}^{(1)}_{V^\prime B^\prime} =&&\int \frac{d^3q}{(2\pi)^3}
\frac{M_{B^\prime}}{E_{B^\prime}} \frac{1}{2\omega_\pi \omega_{V^\prime}}
\frac{1}{P^0-E_{B^\prime}- \omega_{V^\prime} + i\varepsilon}
\frac{1}{-k^0+\omega_{V^\prime}+\omega_\pi- i\varepsilon} \nonumber \\
&&\frac{1}{P^0- \omega_{\pi} - k^0-E_{B^\prime}+ i\varepsilon}
\frac{1}{\omega_\pi+k^0+\omega_{V^\prime}- i\varepsilon} \left[
(\omega_{V^\prime}+\omega_\pi) (\omega_{V^\prime}+\omega_\pi-P^0+E_{B^\prime})+
k^0\omega_\pi\right] \nonumber \\
&&\times \frac{1}{2\vec{k}\,^2} \left\{ \frac{3}{\vec{k}\,^2}
(\vec{q}-\vec{k}\,) \vec{k}\,\,\vec{q}\, \vec{k} -  (\vec{q}-\vec{k}\,)
\vec{q}\right\} F(q) \ ,
\label{eq:g1}
\end{eqnarray}
\begin{eqnarray}
\tilde{G}^{(2)}_{V^\prime B^\prime} =&&\int \frac{d^3q}{(2\pi)^3}
\frac{M_{B^\prime}}{E_{B^\prime}} \frac{1}{2\omega_\pi \omega_{V^\prime}} 
\frac{1}{P^0-E_{B^\prime}- \omega_{V^\prime} + i\varepsilon}
\frac{1}{-k^0+\omega_{V^\prime}+\omega_\pi- i\varepsilon} \nonumber \\
&&\frac{1}{P^0- \omega_{\pi} - k^0-E_{B^\prime}+ i\varepsilon}
\frac{1}{\omega_\pi+k^0+\omega_{V^\prime}- i\varepsilon} \left[
(\omega_{V^\prime}+\omega_\pi) (\omega_{V^\prime}+\omega_\pi-P^0+E_{B^\prime})+
k^0\omega_\pi\right] \nonumber \\
&&\times \frac{1}{2} \left\{ (\vec{q}-\vec{k}\,) \vec{q}-\frac{1}{\vec{k}\,^2}
(\vec{q}-\vec{k}\,) \vec{q}\,\, \vec{q}\, \vec{k}\right\} F(q) \ ,
\label{eq:g2}
\end{eqnarray}
and
\begin{equation}
{G}_{V^\prime B^\prime} =\int \frac{d^3q}{(2\pi)^3}
\frac{M_{B^\prime}}{E_{B^\prime}} \frac{1}{2\omega_{V^\prime}} 
\frac{1}{P^0-E_{B^\prime}- \omega_{V^\prime} + i\varepsilon}\, F(q) \ .
\label{eq:g}
\end{equation}
with
\begin{equation}
P^0=\sqrt{s};~~~\omega_\pi=\sqrt{\vec{q}\,^2+m_\pi^2};~~~\omega_{V^\prime}=\sqrt
{(\vec{q}+\vec{k}\,)^2+m_{V^\prime}^2};
~~~E_{B^\prime}=\sqrt{(\vec{q}+\vec{k}\,)^2+M_{B^\prime}^2} \ .
\end{equation}
We note that the only factor that can produce a pole in the integrand of the former
loop functions
is $(P^0-E_{B^\prime}- \omega_{V^\prime} +
i\varepsilon)^{-1}$ when $P^0$ matches the energy of an intermediate $V^\prime
B^\prime$ state, where $V^\prime=K^*$ and $B^\prime=\Lambda$ or $\Sigma$. In
order to account for the width of
the $K^*$ vector meson, we replace $i\varepsilon$ in this factor by $i
\Gamma/2$, with $\Gamma=50.5$~MeV.

The cross section for the $\gamma N \to K^0 \Sigma$ reactions, obtained after
summing over final spins and averaging over polarizations, including in this way
the contributions of $1/2^-$ and $3/2^-$ states, is given by:
\begin{equation}
\frac{d\sigma_{\gamma N \to K^0 \Sigma}}{d\Omega}=\frac{1}{16\pi^2}\frac{M_N
M_\Sigma}{s}\frac{k}{p} \,\,\overline{\sum}\sum \mid t_{\gamma N \to K^0
\Sigma} \mid^2 \ ,
\label{eq:diff_cross}
\end{equation}
where
\begin{equation}
\overline{\sum}\sum \mid t_{\gamma N \to K^0 \Sigma} \mid^2=\frac{1}{2}\left\{
\left[\mid A \mid^2 \vec{k}\,^2 + 2 \,{\rm Re}\,(A B^*)\right] \vec{k}\,^2\,
\sin^2{\theta} + 2 \mid B \mid^2 \right\} \ .
\label{eq:t2}
\end{equation}

It is obvious from the previous expression that the cross section obtained with
this model produces an angular distribution which is symmetrical with respect to
90$^{\rm o}$ and will not reproduce the forward-peaked cross section
around
$\sqrt{s}=2000$~MeV reported in the experiment \cite{schmieden}. Two mechanisms
were suggested there to play this role: the nucleon-pole term in the s-channel
and the
$\gamma p \to K^0 \Sigma^+$ transition being mediated by $K^*$ exchange in a
t-channel configuration. This second option involves an anomalous $VVP$ coupling,
which is proportional to the relatively large photon momentum and can contribute to the $VB \to V^\prime B^\prime$ transition in the problem considered here. Yet, the term contains a $\vec{\sigma}$ operator from the $PBB$ vertex and does not interfere with the spin-independent transition mediated by vector-meson exchange. Consequently, even if these anomalous contributions might not be negligible in the present problem, they would mostly go to a  background part that would not spoil the essential mechanism for the downfall of the cross section advocated here.  As for implementing a nucleon-pole term, we note that one
should take other resonance contributions as well, especially those that
are close to
the energy region of interest. Given this uncertainty, we have opted for not
adding any supplementary contribution to our model, with the aim of
investigating whether our vector-baryon coupled channel formalism, which
implicitly incorporates the effect of dynamically generated resonances, contains
the basic mechanisms that provide a qualitative explanation of the experimental observations.

\section{Results}

We start this section by checking if the interference effect pointed out
in Ref.~\cite{mishanaka} for the differences between the
$\gamma p \to \eta p$ and $\gamma n \to \eta n$ cross sections in the region of
the $K \Lambda, K \Sigma$ thresholds is also important in the $\gamma N \to
  K^0 \Sigma$ reactions explored in the present work. The authors of
Ref.~\cite{mishanaka} found a destructive interference
between the $K^+ \Lambda$ and $K^+ \Sigma^0$ states excited by the photon
in the $\gamma p \to \eta p$ reaction. Since the $K^+
\Lambda$ is not present in the neutral $\gamma n \to \eta n$ reaction, this
cancellation is absent there, explaining in
this way the enhanced neutral cross section over the charged one. Analogously, in the 
$\gamma N \to
  K^0 \Sigma$ reactions studied here, the photon couples to both the $K^{*
+}\Lambda$ and the $K^{* +}\Sigma^0$ states in the $\gamma p \to K^0 \Sigma^+$
process, but only to the $K^{* +}\Sigma^-$ state in the
case of  $\gamma n \to K^0 \Sigma^0$. This can be easily checked from the 
sum of the direct photon couplings to the $\rho^0$,
$\omega$ and $\phi$ mesons, which implicitly buils up the appropriate
couplings
of the photon to charged vector
mesons.
However, the situation  is slightly more involved in the present work because the amplitude $t_{V N \to V^\prime B^\prime}$, involving rescattering, may induce transitions to states containing a neutral $K^*$ meson, such as $K^{*0}\Sigma^+$ in the $\gamma p \to K^0 \Sigma^+$ reaction and $K^{*0}\Lambda$ in the $\gamma n \to K^0 \Sigma^0$ one (the intermediate $K^{*0}\Sigma^0$ state does not contribute in the latter case because of the zero value of the $\pi^0\Sigma^0\Sigma^0$ coupling at the Yukawa vertex).  As already discussed in the formalism, we are
interested in the region
of the $K^* \Lambda$ threshold, where the anomaly has been seen, and we only
consider  $K^* Y$ states in the intermediate $V^\prime B^\prime$ channels. 
\begin{figure}[h]
\begin{center}
\includegraphics[width=0.6\textwidth]{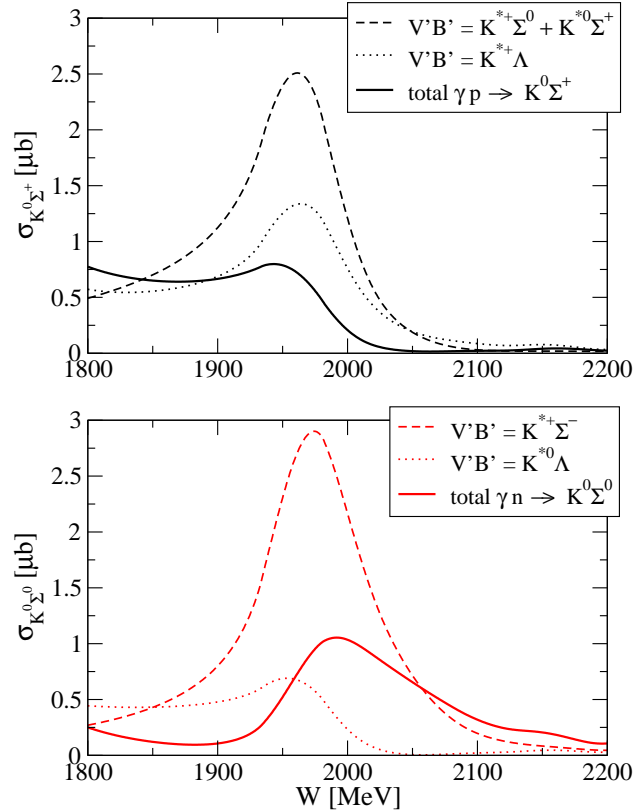}
\caption{Contributions to the $\gamma p \to K^0 \Sigma^+$ (upper panel) and 
$\gamma n \to K^0 \Sigma^0$ (lower panel) cross sections. }
\label{fig:gn_gp}
\end{center}
\end{figure}
In Fig.~\ref{fig:gn_gp} we see how the
consideration of the different intermediate channels, $K^* \Sigma$ (dashed lines)  and $K^* \Lambda$ (dotted lines), build up the final cross section (solid lines), in the
case of the $\gamma p \to K^0 \Sigma^+$ (upper panel) and $\gamma n \to K^0 \Sigma^0$ (lower panel) reactions. As in the work of  Ref.~\cite{mishanaka}, we can 
appreciate in Fig.~\ref{fig:gn_gp}  the tremendous effect of interferences.
The destructive interference between the $K^{*}\Sigma$ and $K^*
\Lambda$ amplitudes, of similar size and shape in the case of  the $\gamma p \to K^0 \Sigma^+$ reaction,
produce an
abrupt downfall of the cross section right at the position of the resonance
generated by the $VB$ interaction model. In contrast, in the
case of the $\gamma n \to K^0 \Sigma^0$ reaction, the 
$K^{*}\Sigma$ and $K^* \Lambda$ amplitudes are quite different, giving rise
to a final cross section retains the peak at the position of
the resonance to a large
extent. In addition, the interference
becomes constructive in between the
$K^*\Lambda$ and $K^*\Sigma$ thresholds, which slows down the decrease of the
neutral cross section at higher energies.

\begin{figure}[h]
\begin{center}
\includegraphics[width=0.5\textwidth]{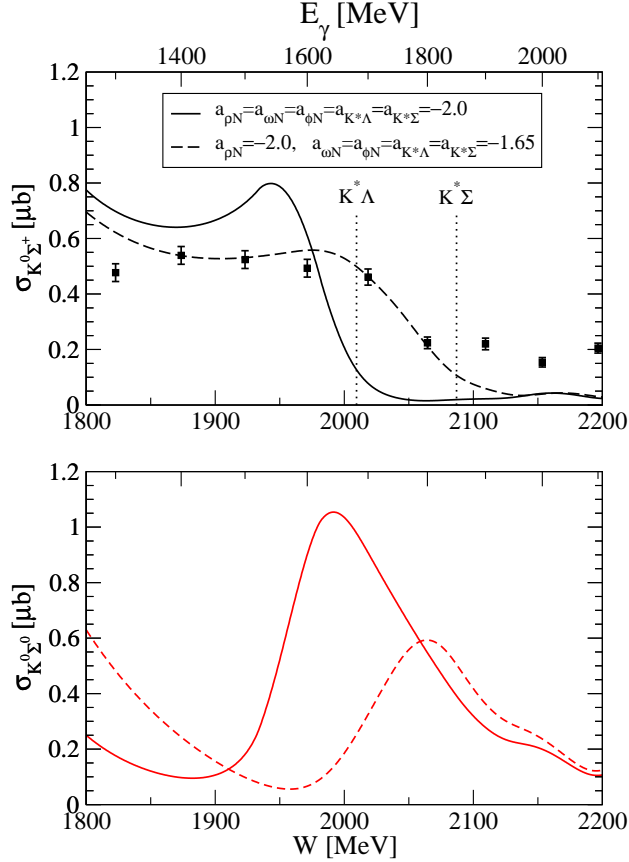}
\caption{Upper panel: Comparison of the $\gamma p \to K^0 \Sigma^+$ cross
section, obtained with two parameter sets, with the CBELSA/TAPS data of
Ref.~\cite{schmieden}. The downfall of the cross section allows one to redefine
the parameters of the model and give a better prediction for the position of the
resonance.
Lower panel: Predictions for the $\gamma n \to K^0 \Sigma^0$ cross section using
two parameter sets. }
\label{fig:sigma}
\end{center}
\end{figure}

We next compare the obtained cross section
for the $\gamma p \to K^0 \Sigma^+$ reaction (solid line) with the
experimental data \cite{schmieden} in
the upper panel of Fig.~\ref{fig:sigma}. 
Note that, as in the experiment, the results in this figure and in all the other ones presented in this work, have been
averaged over the photon energy in bins of $\pm 50$~MeV width. 
We also recall that our model does not intend to obtain a quantitative agreement over
the complete
energy region. In fact, the
behavior at lower $\sqrt{s}$ is not well reproduced. The data has the tendency
to decrease as we lower the energy towards the threshold of the reaction, while our model shows an enhancement close
to $\sqrt{s}=1700$~MeV. This is due to the fact that the employed $VB$
coupled-channel interaction model \cite{angelsvec} produces a narrow
resonance at 1700~MeV, coupling strongly to $\rho N$ and sizably to
$K^*\Lambda$. However, experimentally one finds a much
wider $3/2^-$ resonance \cite{pdg}. In fact,
to reproduce the width of this resonance, which lies below the lowest threshold
of the $VB$ coupled-channel interaction model employed here, it should be
necessary to incorporate, additionally to $\rho N$, lower-lying pseudoscalar-baryon states, such as $\pi N$ in D-wave
and $\pi \Delta$,  as done recently in \cite{Garzon:2013pad}.
However, the incorporation of the $K\Lambda$, $K\Sigma$ channels in the study of the resonance around 2000 MeV has minor effects, as can be seen in Fig.~6 of \cite{javier}.
In view of that, we focus only on interpreting the behavior of the $\gamma p
\to K^0 \Sigma^+$ reaction around the region of the $K^* \Lambda$ threshold.

We
observe that the downfall of the theoretical results displayed by the solid line appears 60~MeV below the
energy at which the experimental cross section presents the abrupt drop.
Actually, since this downfall is sensitive to the position of the resonance
produced in this energy region, we can fine tune the parameters of the model using the CBELSA/TAPS data, 
and obtain in this way a more
realistic prediction of the resonance mass. 
We recall that the vector-baryon interaction model employed \cite{angelsvec}
depends on the subtraction constants $a_l$ ($l=\rho N$, $\omega N$, $\phi N$,
$K^*\Lambda$ and $K^* \Sigma$), used to regularize the meson-baryon loop
function at a regularization scale of $\mu=630$~MeV, that were taken to be have the
natural size value of $-2$, as determined in \cite{Oller:2000fj}.
By changing the subtraction constants to a value of $-1.65$, except 
$a_{\rho N}$ which is kept at its original value of  $-2$,
we obtain the dashed curve in Fig.~\ref{fig:sigma}, which shows the downfall at
the energy
where the experimental cross section presents a sudden drop. This is quite an
achievement of the employed $VB$ coupled-channel interaction, which
explains this behavior from an interference effect between various
resonant amplitudes involving intermediate $K^*\Lambda$ and $K^*\Sigma$ channels, a
feature that none of the isobar phenomenological models, using K-matrix
coupled-channel methods
\cite{maid,said,Anisovich:2005tf,Sarantsev:2005tg,Usov:2005wy}
 but ignoring
$K^* Y$ channels, is able to obtain. Our model could be further tested with
a measurement of the neutral $\gamma n\to K^0 \Sigma^0$ cross section, a prediction of which 
is shown in the lower panel
of Fig.~\ref{fig:sigma} for the
 two different parameter sets employed here. 
%As discussed before, we observe a
%quite different behavior to that of the
%$ \gamma p \to K^0 \Sigma^+$ reaction. 
%In summary, the cross sections for the charged and neutral reactions have a
%completely different energy dependence around 2000~MeV, the later showing a
%peak where the former presents a rapid downfall. We explain this behavior in
%terms of interferences between intermediate $K^*\Lambda$ and $K^*\Sigma$ contributions, embedded
%in the coupled channel
%structure of our model, which are magnified by
%the presence of a nearby resonance.

We can now obtain the characteristics of the
resonance around the energy region of the $K^*\Lambda$ threshold obtained with the
new parameter set. We recall that, in the original model where all the
subtraction constants were set to $-2$ \cite{angelsvec}, the resonance was found to
be represented by a pole in the second Riemann sheet, as
defined e.g. in \cite{Roca:2005nm}, at $z=1977+{\rm i}53$~MeV. It was also pointed out there that, in
a coupled channel model with mesons having a mass distribution, like
$\rho$ and $K^*$ in our case, there is a fuzzy
description of the $VB$  thresholds involving either one of these mesons. When a resonance appears close to
one of such thresholds, finding the pole can be a complicated task. For this reason, in these cases, the resonance
properties were also derived from the shape of the amplitudes in the real energy axis \cite{angelsvec}. In
the case of the resonance discussed here, the real axis method produced a
resonance mass of $M_R=1972$~MeV and a width of $\Gamma_R=64$~MeV. Changing
all the
subtraction constants to $-1.65$, except $a_{\rho N}$ 
which is kept
to its original value of $-2$, the new resonance properties are $M_R=2035$~MeV and
$\Gamma_R=125$~MeV. With the new parameter set the resonance appears 60~MeV up
in  energy, lying now above the $K^* \Lambda$ threshold to which it couples
substantially and, 
consequently,  being almost twice wider. It is interesting to note that
two resonances of negative parity around this region of energy,
$N^*(2080) (3/2^-)$ and $N^*(2090)(1/2^-)$, which appeared in earlier
versions of the PDG, have been eliminated in the latest version \cite{pdg}. A
resonance with these quantum numbers in that energy region, albeit with some
uncertainty in the precise position, is unavoidable in our model due to the attractive
character and strength of the vector-baryon interaction \cite{angelsvec}. These
states also appear in a different work that uses the same
vector-baryon channels, together with the
pseudoscalar-baryon ones and a somewhat different dynamics 
\cite{Gamermann:2011mq}. Experimental support for a $(3/2^-)$ resonance around
2080 MeV has been found recently from an analysis of SPring 8 LEPS data on the
$\gamma p \to K^+ \Lambda(1520)$ reaction in \cite{Xie:2010yk}. Our results
provide another solid backing for the existence of an odd parity resonance
around this energy region. 

\begin{figure}[htb]
\begin{center}
\includegraphics[width=0.8\textwidth]{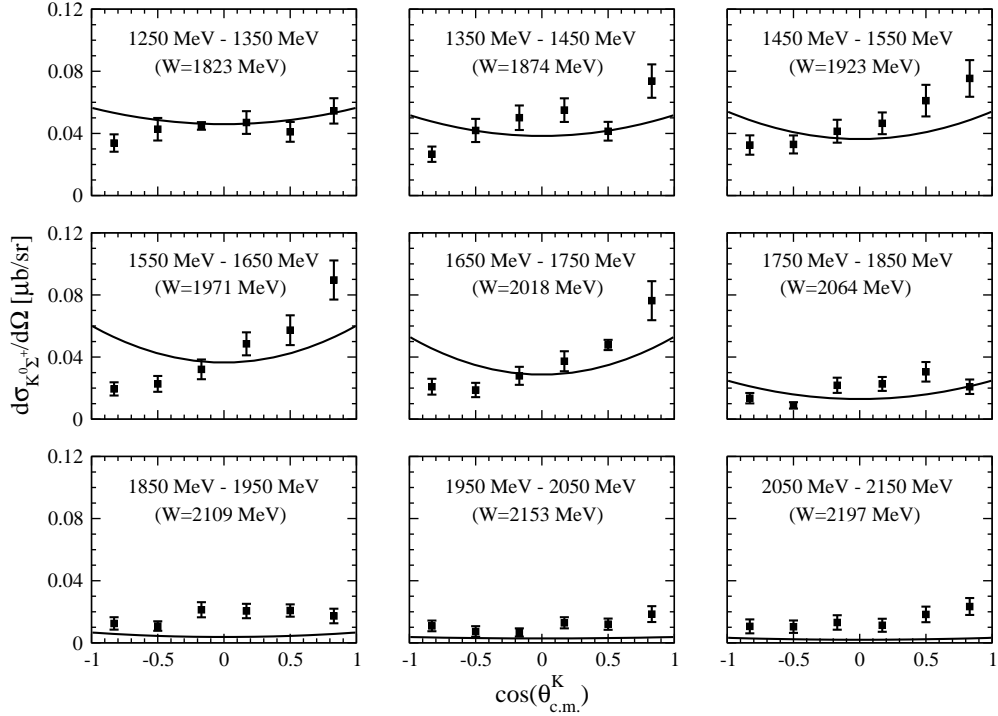}
\caption{Differential cross section for the $\gamma p \to K^0 \Sigma^+$
reaction as a function of the kaon center-of-mass angle, for
several bins of the laboratory photon energy.}
\label{fig:diffp_exp}
\end{center}
\end{figure}

Finally, the calculated differential cross section of the $\gamma p \to K^0
\Sigma^+$ reaction is compared
with the recent experimental CBELSA/TAPS data in Fig.~\ref{fig:diffp_exp}. The
differential cross section is
given as a function of the kaon center-of-mass angle for
several bins of the laboratory photon energy
having a width of $\pm 50$~MeV. The corresponding central center-of-mass $W \equiv \sqrt{s}$
energies
are also quoted in the various panels. 
We already noted that our model predicts a symmetrical angular distribution around 90$^{\rm o}$, which is obvious from the structure of Eq.~(\ref{eq:diff_cross}).
In order to obtain a better agreement with the forward-peaked data, one
would have to extend the
model with P-wave contributions, not-necessarily resonant, that would
interfere constructively at forward angles and destructively at backward ones.
The interesting
thing to realize is that our $VB$ coupled channel model builds up some angular structure with increasing energy, which becomes weaker above the energy where the cross section experiences the downfall, as can be better quantified by the ratio of the difference over the sum of the calculated cross section at $0^{\rm o}$ and $90^{\rm o}$ given in Table~\ref{tab:diff}.
\begin{table}[h]
  \setlength{\tabcolsep}{0.3cm}
\begin{tabular}{|l|c c c c c c c c c|}
\hline
$W$ (MeV) & 1823 & 1874 & 1923 & 1971 & 2018 & 2064 & 2109 & 2153 & 2197 \\
\hline
$\displaystyle\frac{\sigma(0^{\rm o})-\sigma(90^{\rm o})}{\sigma(0^{\rm o})+\sigma(90^{\rm o})}$ &
  0.14 &
0.20 &
 0.24 &
0.29 &
 0.31 &
 0.30 &
 0.23  &
 0.17 &
 0.16\\
\hline
\end{tabular}
\caption{Ratio of the difference over the sum of the calculated $\gamma p \to K^0
\Sigma^+$ cross section at $0^{\rm o}$ and $90^{\rm o}$.}
\label{tab:diff}
\end{table}

\section{Summary and Conclusions}

This work has presented a theoretical study of the $\gamma p \to K^0
\Sigma^+$ reaction around
$\sqrt{ s} = 2000$~MeV, where the cross section shows a rapid downfall and the
differential cross section changes from being forward-peaked to being
essentially flat. The fact that these features
occur between the $K^* \Lambda$ and $K^* \Sigma$
thresholds hints at the plausible explanation that these channels are
crucial ingredients in any theoretical model trying to reproduce the data in
this energy region. In fact, none of the existing K-matrix coupled-channel
resonance models that ignore these channels is able to provide a satisfactory
explanation of the observed structures.

We have developed a model for this reaction that incorporates
the interaction between vector mesons of the nonet with baryons of the octet,
obtained from local hidden gauge
Lagrangians and implementing unitarization in coupled channels. This vector-baryon
interaction produces a resonance in the energy region of
interest that couples strongly to $K^* \Lambda$ and
$K^*\Sigma$ states.

Our results for the  $\gamma p \to K^0
\Sigma^+$ reaction show indeed a rapid drop of the cross section,
which is a consequence of a delicate interference between amplitudes containing
$K^*\Lambda$ and $K^*\Sigma$ intermediate states, magnified by the presence of a resonance that our model produces in this energy region. These interferences are
 present in the  $\gamma n \to K^0
\Sigma^0$ reaction, also studied in this work, although their different balance produces
a neutral cross section that has a maximum where the charged reaction presents the sudden drop.
Since this structure is sensitive to the position
of the resonance, we have
used the CBELSA/TAPS data to fine tune the parameters of our model and obtain in
this way a more realistic prediction of the resonance mass, that we now find at
2030~MeV, in the energy region where earlier versions of the PDG pointed out the
presence of two close resonances with spin-parity $1/2^-$ and $3/2^-$. The
important role played by a resonant $N^*(2080)$ contribution has also been
pointed out recently from an analysis of SPring8 LEPS $K^+\Lambda(1520)$
photoproduction data. A measurement of the
interesting signature that this resonance leaves on the $\gamma n \to K^0
\Sigma^0$ reaction, showing a peak structure
where the $\gamma p \to K^0 \Sigma^+$ reaction has a sharp downfall, would give
extra support to the existence of this state.

We have also obtained the $\gamma p \to K^0
\Sigma^+$ differential cross sections and have observed the transition to a
weaker
angular distribution above the energy where the cross section experiences the downfall, as in experiment.

The model presented here is by no means a complete one. Explicit background
contributions and genuine resonant terms would have to be included to
reproduce the complete set of data, from the $K^0
\Sigma$ threshold up to the energy region explored here. However, this study has
brought up the important observation that the consideration of the $K^*
\Lambda$ and $K^*\Sigma$ states in a unitary
coupled-channel scheme is crucial to reproduce the rapid
downfall of the $\gamma p \to K^0 \Sigma^+$ cross section around  
2000~MeV. This
hypothesis would be further tested by a measurement of the neutral reaction
$\gamma n \to K^0\Sigma^0$, which can bring further light
into the physics hidden in these processes.

\section*{Acknowledgments}  
This work is partly supported by the Spanish Ministerio de Economia y
Competitividad and European FEDER funds under the contract numbers
FIS2011-28853-C02-01 and FIS2011-24154, by the Generalitat Valenciana in
the program Prometeo 2009/090, and by Grant
No. 2009SGR-1289 from the Generalitat de Catalunya. We also acknowledge
the support of the Consolider
Ingenio 2010 Programme CPAN CSD2007-00042 and of the European
Community-Research Infrastructure Integrating Activity
Study of Strongly Interacting Matter (acronym HadronPhysics3, Grant Agreement
n. 283286) under the Seventh Framework Programme of EU.

\end{document}